\documentclass[twocolumn,floats,aps,showpacs]{revtex4}
\usepackage{times,amsmath,graphics,psfrag,latexsym}
\usepackage{pstricks}

\begin{document}  
\title{The Influence of Shape on Ordering of Granular Systems 
in Two Dimensions} 
\author{I.C. Rankenburg and R.J. Zieve}
\address{Physics Department, University of California at Davis}  
\begin{abstract}  
We investigate ordering properties of two-dimensional granular materials using
several shapes created by welding ball bearings together.   Ordered domains
form much more easily in two than in three dimensions, even when configurations
lack long-range order. The onset of ordered domains occurs near a packing
density of 0.8, a phenomenon observed previously for disks.  One of our shapes,
the trapezoid, has packings that remain disordered and near the transition
density even after annealing by shaking.  Since random packings are unstable
for disks and many other shapes in two dimensions, trapezoid packings provide
a new approach to studying two-dimensional randomness. We also find that the
rotational symmetry of a shape is an excellent predictor of how easily it
orders, and a potential guide to identifying two-dimensional shapes that remain
random after annealing.  
\end{abstract}
\pacs{45.70Cc, 81.05 Rm}
\maketitle

\section{Introduction}

Granular materials are studied in materials science, geology, physics, and
engineering.  Their unusual dynamical behavior has attracted much recent
attention, but even their most basic static properties are not yet understood
\cite{German, Cumberland}. For example, a collection of uniform spheres in
three dimensions reaches a final arrangement known as random close packing
(RCP).  The RCP density is 0.64, substantially smaller than the 0.74 density of
the fcc and hcp lattices.  The RCP density is robust to changes in the
vibration method, or to using uniaxial or hydrostatic pressure to push the
grains together \cite{3Dexpt}. Computer simulations also agree well with
physical experiments \cite{simJodrey}. The structure of RCP arrangements is
thoroughly documented by experiments and simulations, but neither the
arrangements nor their density is understood from fundamental principles.
The only algorithms for generating an RCP configuration involve a
simulated shaking procedure. 
Even attempts to derive the RCP density begin from experimentally measured 
correlation functions \cite{Berryman}.

As a further complication, 
direct applications of packing densities often deal with non-spherical
particles, or mixtures of shapes or sizes.  In mixing dyes, the
highest possible percentage of colorant depends on the RCP density
for the molecule's shape \cite{dyes}.  In making ceramics, particles
form clusters whose shape affects the final density \cite{ceramics}.
Particle shape influences packing densities in complicated ways. The
ideal maximum packing density, as well as the RCP density, depends on
shape and is unknown in most cases.  Furthermore, the density observed
in a given experiment also depends on how easily particles move past
each other into optimal positions. Interparticle friction can be large
when a shape has flat sides, and multiple contacts can eliminate much
of the particles' rotational freedom. Cubes, for example, have maximum
packing density of 1 but reach a density of only 0.68 on deposition
\cite{cubes}. On the other hand, irregularly shaped particles are less
dense than spheres on initial deposition but compress particularly well
with vibration \cite{shapedep}.

Theoretical work has concentrated on packings of hard spheres
\cite{simJodrey, Berryman}, with generalizations to hard ellipsoids
\cite{ellipseCuesta, ellipseVB, ellipseBuchalter,ellipseSherwood} or
mixtures of sphere sizes \cite{simKausch, simClarke}. For ellipsoids,
simulations find several phases as a function of particle density.
Theory has rarely dealt with other shapes, even ``simple" ones such as
regular polygons and polyhedra.  Without understanding the mechanisms
by which such particles move into position, the utility of simulations
is unclear.

For many physics issues, solving an analogous problem in two dimensions can
provide insight to the full three-dimensional question.  However, there is no
stable random configuration of circles in two dimensions.  Uniform spheres
confined to a single layer easily form a triangular lattice, the densest
possible packing.  Experiments and simulations do suggest a transition between
random and ordered configurations near a density of 0.80 \cite{Turnbull, QT,
Kierlik}. 
Density increases much more slowly beyond this point.  The number of touching
disks in the configuration also changes sharply with density near 0.80. 
However, without a better  definition of a ``random" arrangement, pinpointing
the exact transition  and analyzing two-dimensional random close-packed
configurations is impossible. 

Not only spheres, but even unusual shapes such as regular pentagons anneal to 
their densest known packing in two dimensions \cite{Troadec}.  The arrangement
for pentagons  is a double lattice, in which translates of a two-pentagon unit
cover the plane.  Computer simulations find an analogous double lattice for
heptagons as well \cite{heptagon}.  Thus these shapes too are impractical for
studying random close packed structures.

Here we study several shapes in two dimensions.  A main goal is to find shapes
that remain disordered, which would lend themselves to studies of random
arrangements.  In addition, we may better understand the dependence of ordering
properties on shape.  Finally, packing in two dimensions is important in its
own right in the behavior of films and monolayers.

All our shapes are clusters of spheres welded into triangular lattice
positions.   This guarantees that the densest packing is always a triangular
lattice of the component spheres.  The point contacts between clusters minimize
friction and  blocking effects as the shapes move past each other.  Another
advantage is that comparison with computer experiments is possible, since
overlaps are easy to check with spheres.  Visualizing the  entire arrangement
is also far easier in two dimensions than in three.

\begin{figure}[hbt]
\scalebox{.32}{\includegraphics{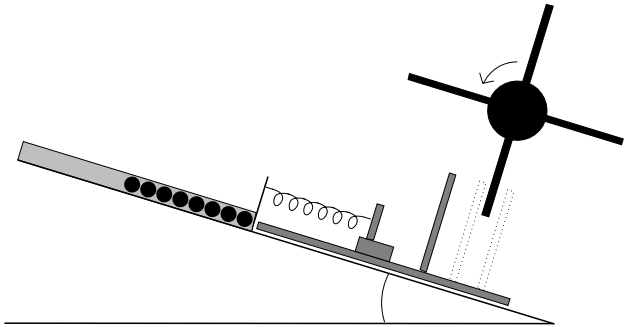}}
\caption{A schematic of the experiment.  Moving the arm of the sliding piece
to one of the outlined sites changes the spring tension at the moment of
release and hence the hit magnitude.}
\label{f:twacker}
\end{figure}

\section{Procedure}

We welded together $\frac18$-inch diameter carbon steel ball bearings with a
Unitek 60 welder set at its maximum power, 60 Watts.  During the welding  the
balls were held in a bakelite mount machined to give the desired final shape. 
The shapes made were doubles, triples (three balls in a straight line),
triangles of three or six  balls, diamonds of four balls, trapezoids of five
balls, and hexagons of seven balls. The welding appears not to distort the
balls.  One of the most stringent tests is that, as we shall see, several of
the welded shapes order into perfect triangular lattices of their individual
balls. A significant distortion from the welding would destroy the long-range
order.

A schematic of our shaking apparatus is shown in Figure \ref{f:twacker}.   Two
pieces of plexiglas separated by 0.135-inch spacers confine balls inside to a
single layer.  The container is placed at an angle to the horizontal so that
gravity pulls the balls towards one side.  We typically fill a 9 inch by 4 inch
region with ball clusters.  Roughly 2500 single balls fit in this space.  After
putting the shapes in, we shake the box roughly to create a disordered initial
state. An aluminum plate, connected to the  plexiglas by a spring, serves as a
hammer.  Rotating spokes pull back and release the aluminum plate.  On release,
it strikes the bottom of the container and shakes the balls. A hit occurs once
every three seconds for one hour.  The balls stop moving completely between
hits.  The configuration changes substantially during the first few shakes, and
negligibly at the end of the hour, but we have not investigated quantitatively
the rate of ordering.  

\begin{figure}[thb]
\scalebox{.6}{\includegraphics{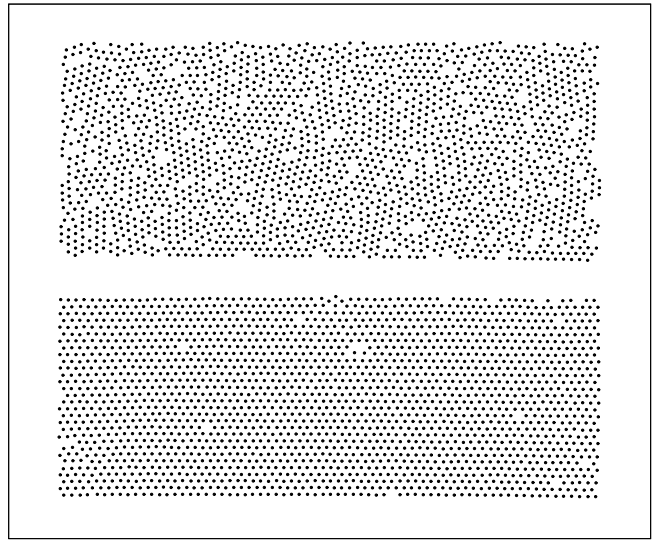}}
\caption{The centers of the balls, as identified by the computer from
digital photographs of a trapezoid initial condition (top) and a well
ordered arrangement of doubles (bottom). }
\label{f:goodbadpic}
\end{figure}

We vary both the maximum spring extension and the angle of the container.
These parameters change the relative strengths of three physical 
forces: the hit magnitude, the component of gravity pulling the balls together,
and the friction force between the balls and the plexiglas.  Friction and 
effective gravity depend only on the angle, with friction decreasing as the
angle becomes steeper and gravity increasing.  We use eighteen different
settings involving three spring extensions and seven angles between 20$^\circ$ 
and 50$^\circ$.

The hit magnitude depends on both spring length and angle, since the weight of
the striking plate itself changes the equilibrium length of the spring by an
angle-dependent amount. A very hard hit destroys the memory of the situation,
and each time the balls settle from scratch.  Furthermore, very strong hits can
break the weld joints.  On the other hand, a hit so light that the displacement
of the balls is small compared to the ball diameter makes rearrangement
difficult and slow.  In addition, for light hits at the smallest angles shapes
occasionally stop moving in positions that clearly should be unstable, 
held in place only by friction.  Thus we might expect an optimum hit magnitude,
not necessarily the same for different shapes.  Using a range of parameters
lets us determine not only whether but also how easily different shapes order.

\begin{figure}[tbh]
\scalebox{.68}{\includegraphics{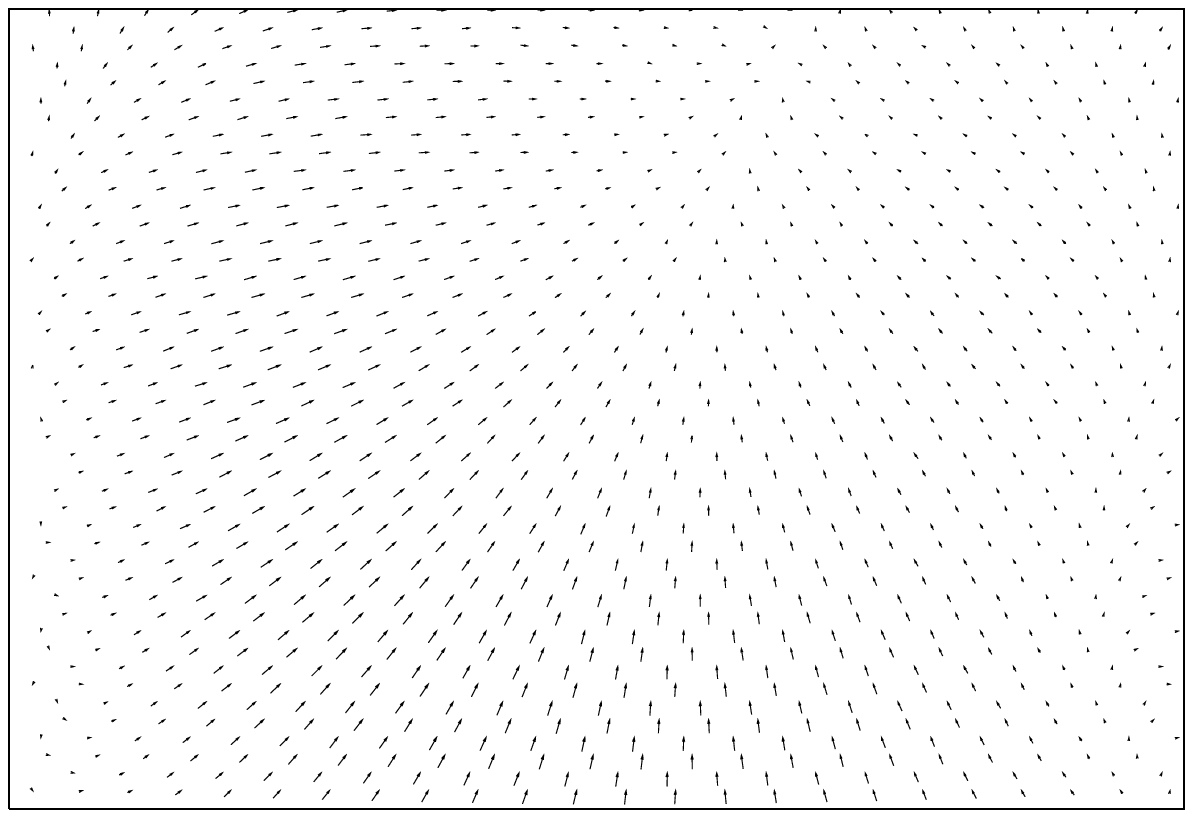}}
\caption{This vector field shows the displacements from true positions due
to the photographs.  The entire region is roughly $70 \times 30$ ball diameters.}
\label{f:warping}
\end{figure}

\begin{figure*}[tbh]
\includegraphics{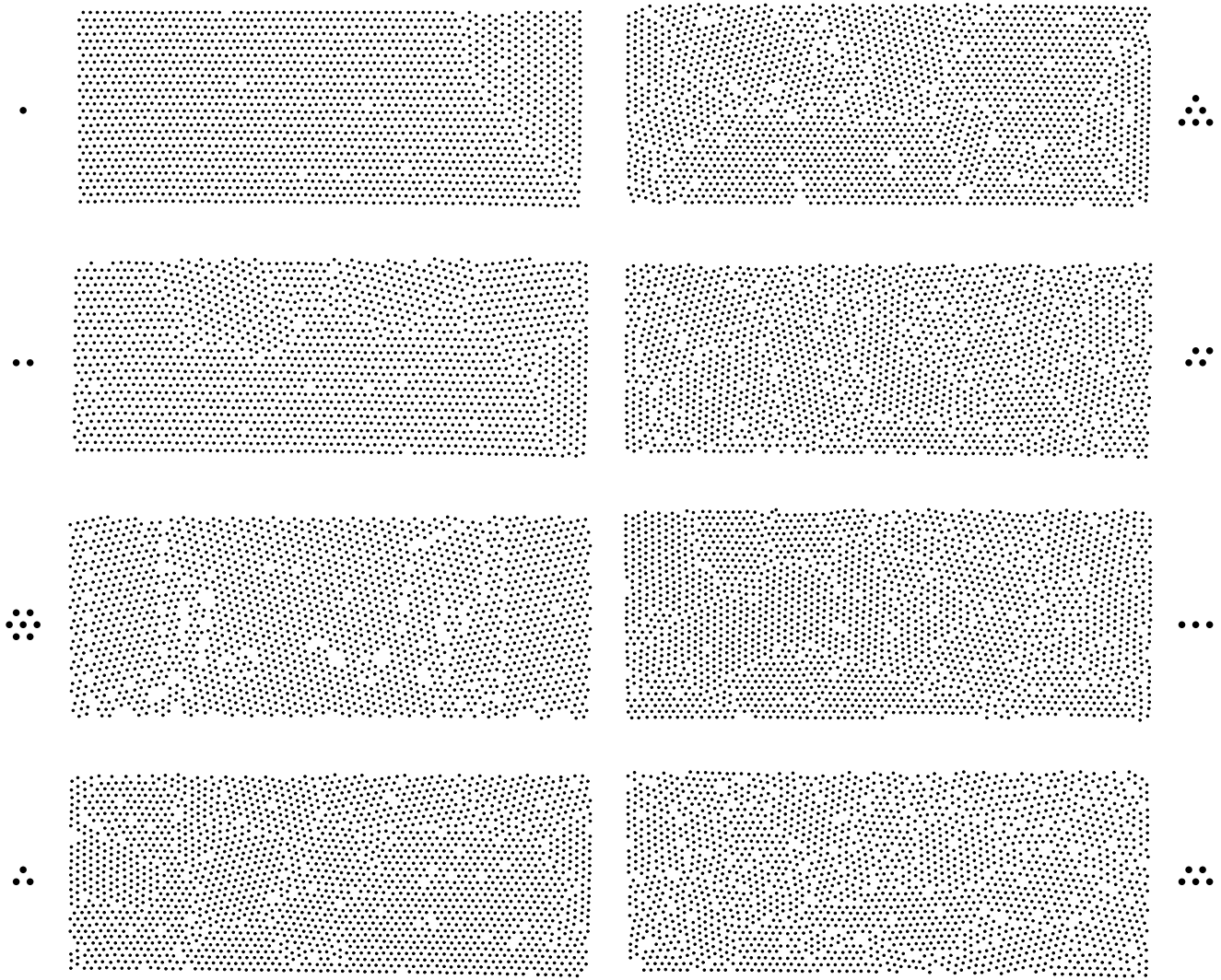}
\caption{Final configurations for different shapes for an angle of $40^{\circ}$
and intermediate spring setting.  Left, top to bottom: singles,
doubles, hexagons, small triangles.  Right, top to bottom: large triangles,
diamonds, triples, trapezoids.}
\label{f:endconfigs}
\end{figure*}

When the shaking finishes, we photograph the final configuration with an
Olympus 340R digital camera and transfer the image to a computer.  The computer
scans the photograph for local bright spots, which appear near the center of
each ball.  We refine the positions with a weighted average of the pixels
surrounding the bright spots.  Figure \ref{f:goodbadpic} shows the computer's
identification of an initial state and a well-ordered final state.  To locate
the balls successfully by this method, a single bright light source is used
when taking the photograph.  Multiple sources can lead to several bright spots
on a single ball, or spots far from the ball center.  A significant systematic
error comes from warping of the photograph by the camera lens itself.  To
eliminate the warping, we generate a perfect triangular lattice of black dots
on white paper and photograph it.   We then find the displacement from ideal
lattice sites of each photographed  dot and interpolate to get the displacement
at other points in the picture. Figure \ref{f:warping} shows the resulting
vector field of displacements.  We adjust the position of each ball center in
the  photographs to compensate for the warping.  The maximum correction is 
comparable to one ball diameter.

\begin{table*}[t]
\caption{Highest and lowest density for different shapes.  The numbers in
parentheses show what the densities would be if the holes were filled in.} 
\label{t:density}
\newcommand{\m}{\hphantom{$-$}} 
\newcommand{\cc}[1]{\multicolumn{1}{c}{#1}}
\renewcommand{\tabcolsep}{1pc} 
\renewcommand{\arraystretch}{1.2} 
\begin{tabular}{@{}|l|ccc|ccc|c|} 
\hline &  & Highest &  & & Lowest& & Initial\\
\hline 
Doubles         & 0.908  & 0.908  & 0.905  & 0.876 & 0.870  & 0.865  & 0.825\\
  & (0.912) & (0.911) & (0.909) & (0.881) & (0.875) & (0.873) & \\
\hline 
Hexagons        & 0.887 & 0.886  & 0.878  & 0.833 & 0.832  & 0.801 & 0.784\\
     & (0.895) & (0.894) &(0.891) &  (0.861) &  (0.850) & (0.829) & \\
\hline 
Small Triangles & 0.881 & 0.879  & 0.877  & 0.848 & 0.845 & 0.844 & 0.820\\
 & (0.892) & (0.889) &(0.887) &  (0.861) &  (0.855) & (0.853) & \\
\hline 
Large Triangles & 0.863 & 0.861  & 0.854  & 0.830 & 0.829  & 0.827 & 0.813\\
 & (0.883) & (0.880) &(0.872) &  (0.842) & (0.839) & (0.837) & \\
\hline 
Diamonds        & 0.853 & 0.852  & 0.851  & 0.820 & 0.814  & 0.808 & 0.793\\
    & (0.880) & (0.870) &(0.869) &  (0.832) & (0.822) & (0.814) & \\
\hline 
Triples         & 0.862 & 0.861  & 0.859  & 0.835 & 0.816  & 0.813 & 0.792\\
    & (0.874) & (0.873) &(0.870) &  (0.841) & (0.823) & (0.816) & \\
\hline 
Trapezoids      & 0.839 & 0.836  & 0.832  & 0.812 & 0.809  & 0.808 & 0.788\\
    & (0.864) & (0.853) &(0.852) &  (0.826) &  (0.824) & (0.819) & \\
\hline
\end{tabular} 
\end{table*}

After each trial we sort the shapes and verify that less than 7\% of the shapes
broke.  The amount of breakage correlates strongly to both hit strength and
shape.  There is usually no breakage until mid-level hits and less than 3\% on
all but the hardest few settings.  Shapes such as 3-ball triangles, with each
ball welded to two neighbors, are strong.  Triples, with two balls held by a
single weld, break much more easily.  Breakage seems to have little effect on
the ordering.  For all shapes, the run with the most breakage was not the one
with the best ordering.  Even for triples, where broken shapes become singles
and doubles, which both order easily, high breakage did not correlate to good
order.  A few settings for triples and 6-ball triangles, and one for doubles,
were omitted because of large amounts of breakage.

\section{Results}

As we vary the shaking parameters, all shapes except trapezoids and triples
form domains comparable in size to the container itself, but they do so under
increasingly restrictive hit conditions.  Figure \ref{f:endconfigs}, which
shows final configurations for parameters away from optimum, reveals the
substantial differences among shapes and illustrates many of the trends we
find.  The arrangements all come from a 40$^\circ$ angle and the middle of the
three spring settings.  Singles, doubles and hexagons can form essentially 
perfect lattices, with a few large domains oriented by the container walls. For
triangles and diamonds, small domains always appear in addition to the large
ones.  Domain sizes for triples and trapezoids are always substantially smaller
than the container size, suggesting an absence of long-range order.  Usually
the best ordering occurs at an intermediate hit power.  Trapezoids are an
exception, with the degree of order nearly independent of hit settings.

In all the arrangements of Figure \ref{f:endconfigs}, the individual balls
composing the shapes form ordered domains with sharp boundaries.   The
configurations have progressively decreasing domain sizes.  As the domains
shrink, interstitial clusters and regions that appear ``random" increase in
size and number. Much of the configuration for trapezoids, the least ordered
shape, has ordered regions no more than 6 balls (or two trapezoids) across. 
This small domain size suggests that trapezoids approach a regime of stable
random arrangements.  We stress that our notion of ``order" involves only the
arrangement of the individual spheres.  Identifying which balls are welded
together is difficult.  We have done so for a few cases and find
no long-range orientational order beyond the requirement that the individual
balls lie in a lattice.  Thus we are really studying the packing of
disks in a plane subject to certain constraints.

The shapes also differ in the voids that appear.  Triples give rise to fewer
holes than do the non-linear shapes.  Larger shapes support larger
holes, sometimes with characteristic shapes.  For example, 
hexagon packings regularly show hexagonal holes, and
the large (6-ball) triangles often have three balls missing in a triangular
pattern.  Hexagons frequently have rows of holes, as on the far left of
Figure \ref{f:endconfigs}, as well as substantial void regions at crystallite
boundaries.

\section{Large-scale order}

We quantify the degree of order in several ways.  Density, shown in Table
\ref{t:density}, measures the state of the entire system.  
We use a large rectangle and calculate the total ball area inside,
contributed both by balls completely inside and by balls on the boundary.
Our biggest source of error in the density calculation comes from the
lengths of the rectangle's sides in units of ball diameters.  

The densities confirm the qualitative discussion on the relative ordering of
the different shapes in Figure \ref{f:endconfigs}.  Of the two main defects in
our arrangements, holes and grain boundaries, holes have a much larger effect
on density, particularly for the largest shapes.  Revised densities, treating
holes in an otherwise crystalline region as filled, are shown in parentheses 
in Table \ref{t:density}.  Small triangles pack slightly better than large
triangles, although the difference decreases when holes are counted as filled. 
The similar behavior of the two sizes of triangles, particularly after
discounting holes, shows that the ordering behavior is not completely dominated
by size-dependent effects such as the container size or hit power.  

The density of a perfect triangular lattice, and the maximum possible
density for all of our shapes, is 0.9069.  (That the densest packings
for doubles in Table \ref{t:density} apparently exceed this value is merely
a consequence of the uncertainty in the density calculation.  There
may be a systematic error of less than 0.5\% toward larger density, perhaps
because the container height is slightly larger than the ball diameter
and the balls may not lie perfectly flat.) For
packing of spheres in two dimensions, a transition from random to
ordered structures occurs near 0.80. In one experiment a gradual,
uniform contraction of a rubber sheet increases the density of disks
lying on the sheet \cite{QT}.  After the disks come into contact, they
slide across the sheet under further contraction.  The disk motion from
sliding causes hexagonal crystals to form at a density of about 0.82.
Another technique \cite{Turnbull} shakes a fixed-density system to find
the most probable configurations.  Repeating with a series of densities
shows a change in behavior near 0.75, with large lattice regions forming
above this density.  Computer simulations of a liquid-solid interface have
put a hard-sphere liquid near a two-dimensional lattice of attractive
sites and find a possible first-order transition due to packing effects
at a density 0.72 \cite{Kierlik}.  The restriction to lattice sites
naturally reduces the transition density.

\begin{figure}[b]
\scalebox{.55}{\includegraphics{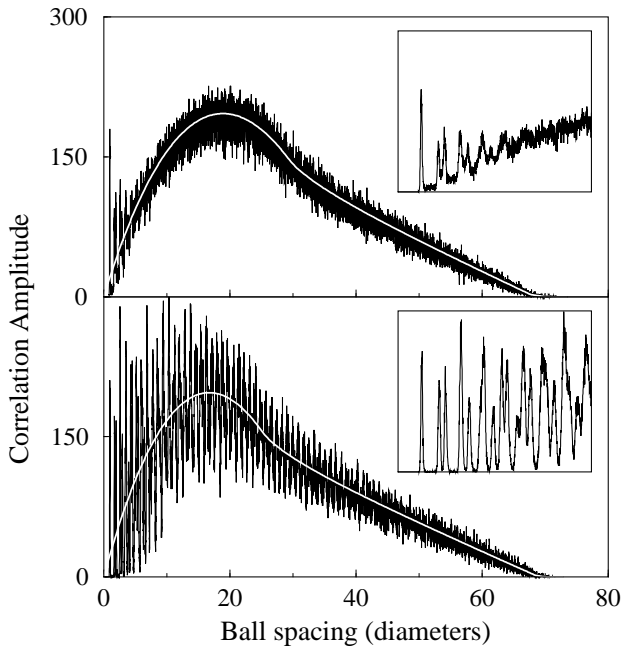}}
\caption{Pair correlation functions corresponding to the two-ball
configurations of Figure \ref{f:goodbadpic}.  The white lines are average
values calculated from Eqs. (1)-(3), as described in the text.  The insets
expand the regions from zero to eight ball diameters.}
\label{f:goodbadhist}
\end{figure}

Interestingly, our lowest densities, found for trapezoids, are very close to
the transition density, and qualitatively there appear to be  only small
domains in these packings.  The agreement of the trapezoid density with the
previously observed hard-disk transition density supports the idea of using
final trapezoid arrangements as models of two-dimensional RCP configurations. 
Although the exact configurations attained by trapezoids would be unstable as
arrangements of unwelded single balls, the maximum random density should be
similar for the two cases.  The perfect alignment of the balls within each
trapezoid increases the maximum random density, but the larger holes in
trapezoid packings decrease the density. 

The comparison with the less well-ordered configurations also illustrates how
easily different shapes order.  Even with very light taps, doubles order
reasonably well, for instance.  Not surprisingly, the degree of order under
light taps has strong correlation to the size of the shapes used.

To identify a length scale for the domains in each arrangement, we
calculate a two-ball correlation function.  We 
consider the distance between each pair of balls and construct a histogram of
these distances, with about 320 bins per ball diameter.  This is equivalent 
to finding the number of ball centers in
annuli of fixed width centered at a single ball. For a perfect lattice, the
result is sharp peaks at spacings characteristic of the triangular lattice:
1,$\sqrt{3}$, 2, etc. diameters.  Imperfections, both in the identification of
the ball centers and in the lattice itself, broaden the peaks and raise the
heights of the intervening values.  Figure \ref{f:goodbadhist} shows the
histograms corresponding to the lattices of Figure \ref{f:goodbadpic}.   

Geometrical considerations determine the average shape of the histogram.
The linear increase in average amplitude with distance at small separations
corresponds to the increase in the circumference of a circle with its diameter. 
The average amplitude in the histogram falls off at larger distances because of
the finite size of our sample.  We calculate the exact distribution of 
separations for two balls randomly positioned in a rectangle, as follows.

\begin{figure}[b]
\includegraphics{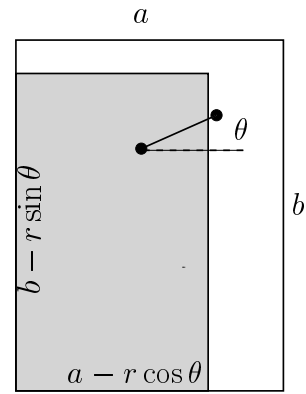}
\caption{Geometry of the average histogram calculation described in the text. 
If two points in the larger rectangle have the angle and separation shown, the
lower left point always lies within the shaded region.}
\label{f:avgcalc}
\end{figure}

\begin{table*}[tb]
\caption{Extent of order, in ball diameters.  The length scale of the 
most ordered arrangements is limited by the sample size.}
\label{t:longrangeorder}
\newcommand{\m}{\hphantom{$-$}} 
\newcommand{\cc}[1]{\multicolumn{1}{c}{#1}}
\renewcommand{\tabcolsep}{1pc} 
\renewcommand{\arraystretch}{1.2} 
\begin{tabular}{@{}l|ccc|ccc|c} 
\hline &  & Most ordered &  & & Least ordered & & Initial\\
\hline 
Doubles         & 30.2 & 27.7 & 27.0 & 14.3& 13.0& 10.2& 3.0\\
Hexagons        & 28.0 & 25.5 & 25.1 & 7.8 & 7.6 & 7.5 & 3.3\\
Small Triangles & 25.3 & 22.1 & 20.4 & 8.1 & 8.0 & 7.7 & 2.7\\
Large Triangles & 20.6 & 14.5 & 13.8 & 7.9 & 6.8 & 6.8 & 2.7\\
       Diamonds & 13.6 & 13.1 & 11.7 & 4.6 & 4.6 & 3.7 & 3.0\\
        Triples & 12.2 & 10.2 & 8.1  & 4.5 & 3.0 & 2.0 & 2.0\\
     Trapezoids & 6.0  & 6.0  & 5.6  & 4.3 & 3.7 & 3.6 & 2.5\\
\hline
\end{tabular} 
\end{table*}

\begin{table*}[tb]
\caption{$D$, a weighted ratio of the first few peak and valley amplitudes 
in the two-ball correlation function.  See text.} 
\label{t:peakratio}
\newcommand{\m}{\hphantom{$-$}} 
\newcommand{\cc}[1]{\multicolumn{1}{c}{#1}}
\renewcommand{\tabcolsep}{1pc} 
\renewcommand{\arraystretch}{1.2} 
\begin{tabular}{@{}l|ccc|ccc|c} 
\hline &  & Highest &  & & Lowest & & Initial\\
\hline 
Doubles         & 1480  & 1451  & 642.6 & 78.7 & 76.2 & 64.2 & 16.0\\
Hexagons        & 410.8 & 216.8 & 191.0 & 54.2 & 42.3 & 33.6 & 26.7\\
Small Triangles & 109.2 & 105.3 & 104.7 & 43.1 & 42.9 & 40.8 & 20.5\\
Large Triangles & 91.8  & 88.6  & 82.4  & 44.2 & 38.9 & 35.0 & 25.3\\
       Diamonds & 71.0  & 68.7  & 61.8  & 26.5 & 23.9 & 20.3 & 18.6\\
        Triples & 57.6  & 50.5  & 44.4  & 23.6 & 17.9 & 16.2 & 14.1\\
     Trapezoids & 43.1  & 35.6  & 32.9  & 24.7 & 24.5 & 21.5 & 20.1\\
\hline
\end{tabular} 
\end{table*}

Let $a$ and $b$ be the lengths of the rectangle's sides, with $a\leq b$.
The region of phase space corresponding to choosing two points at random in the
rectangle has volume $(ab)^2$. Now calculate the volume of the subset with
point separation between $r$ and $r+dr$.  With the rectangle oriented as in
Figure \ref{f:avgcalc}, the first point lies below and to the left of the
second point  one quarter of the time.  Let $\theta$ be the angle between the
line connecting the points and the rectangle's short side. If $r\leq a\leq b$,
the first ball can lie anywhere inside the shaded rectangle, of sides
$a-r\cos\theta$ and $b-r\sin\theta$.  Integrating over $\theta$, the volume
corresponding to separation $r$ to $r+dr$ is 
\begin{eqnarray}
rdr & \int_0^{\pi/2} d\theta 
4(a-r\cos\theta)(b-r\sin\theta)&\\ & =[2\pi ab -4(a+b)r+2r^2]rdr. &\nonumber
\end{eqnarray}
For $a\leq r\leq b$, the same integral applies, except that not all angles 
are allowed.  The result is 
\begin{align}
4r&dr \int_{\cos^{-1}\frac ar}^{\pi/2} d\theta(a-r\cos\theta)(b-r\sin\theta)=
\\
&[2\pi ab-4ab\cos^{-1}\frac ar -2a^2-4b(r-\sqrt{r^2-a^2})]rdr. \nonumber
\end{align}

Finally, for $a\leq b\leq r\leq \sqrt{a^2+b^2}$, there is an additional
restriction on the allowed $\theta$, giving 
\begin{align}
4rdr&\int_{\cos^{-1}\frac ar}^{\sin^{-1}\frac br} d\theta(a-r\cos\theta)
(b-r\sin\theta)=\\ \nonumber &[4ab(\sin^{-1}\frac br-\cos^{-1}\frac
ar)+4a\sqrt{r^2- a^2} \\ & +4b\sqrt{r^2-a^2} -2(a^2+b^2+r^2)]rdr. \nonumber
\end{align}

These formulas, divided by $(ab)^2$, give the probability that two points are
separated by a distance between $r$ and $r+dr$.  For a region containing $N$
balls, the average value of the pair correlation function is this function
times the number of pairs, $N\choose 2$. The white lines of Figure 
\ref{f:goodbadhist} are the average values calculated in this way.

The two parts of Figure \ref{f:goodbadhist} differ most obviously in the extent
of the structure.  In the upper graph, peaks give way to noise near 4 ball
diameters, whereas the lower graph retains structure to 40 diameters. To
measure the long-range correlations, we begin by taking the difference between
the observed histogram and its calculated average value.   We average the 50
highest values within a window 2 ball diameters wide, and also the 50
lowest values.  Then we take the difference between the two. As the
window moves to larger distances, this difference generally decreases.
It approaches 50 for the largest distances, with extremely disordered
lattices reaching this value within a few ball diameters. To assign
a length scale to a configuration, we set a cutoff of 80.  The best
lattices drop below 80 near 30 ball diameters, which is comparable to the
system size in one direction.  Even primarily crystalline arrangements
often have two or three domains, aligned with the different walls of the
container, and the length scale we find depends on the relative sizes
of these domains. The length scales for several different configurations
are shown in Table \ref{t:longrangeorder}.

\section{Local order}

To further characterize the appearance of holes and grain boundaries in the
different shapes, we analyze short-range order using methods sensitive to each
type of imperfection.  Domain boundaries strongly influence the short-distance
region of the two-ball correlation function, while coordination numbers are
more sensitive to voids.

\begin{table*}[t]
\caption{Percentages of balls with hexagonal Voronoi regions,
for several shapes.  The highest and lowest percentages among 
final packings, as well as the initial percentage, are shown.} 
\label{t:Voronoi}
\newcommand{\m}{\hphantom{$-$}} 
\newcommand{\cc}[1]{\multicolumn{1}{c}{#1}}
\renewcommand{\tabcolsep}{1pc} 
\renewcommand{\arraystretch}{1.2} 
\begin{tabular}{@{}l|ccc|ccc|c} 
\hline &  & Highest &  & & Lowest & & Initial\\
\hline Doubles & 98.7 & 98.2 & 97.6 & 88.2 & 87.1 & 86.8 & 74.4\\
Hexagons &  96.2 & 94.1 & 93.3 & 80.8 & 80.5 & 73.3 & 70.8\\
Small Triangles & 89.9 & 89.5 & 88.6 & 82.1 & 82.0 & 82.0 & 73.1\\
Large Triangles & 88.0 & 87.8 & 87.6 & 83.3 & 83.1 & 80.6 & 78.4\\
Diamonds & 87.3 & 85.3 & 83.8 & 73.9 & 73.0 & 71.3 & 69.0\\
Triples & 86.2 & 85.7 & 85.7 & 77.8 & 72.8 & 71.3 & 67.3\\
Trapezoids & 78.1 & 78.1 & 77.6 & 71.7 & 71.5 & 70.8 & 67.4\\
\hline
\end{tabular} 
\end{table*}
   
We use the correlation function out to 3 ball diameters.  We add the
heights of the first five peaks, and divide by the sum of the first and third
valley heights.  We omit the second and fourth valleys because they lie between
closely spaced peaks and give less consistent results.  This quotient $D$ is 
larger for better ordered configurations. For perfect order, no balls lie in
the valleys and the ratio is infinite.  For the most disordered system we can
engineer, the value is about 20.  Using a ratio means that the exact number of
balls used for each shape is unimportant.

\begin{figure}[b]
\scalebox{.45}{\includegraphics{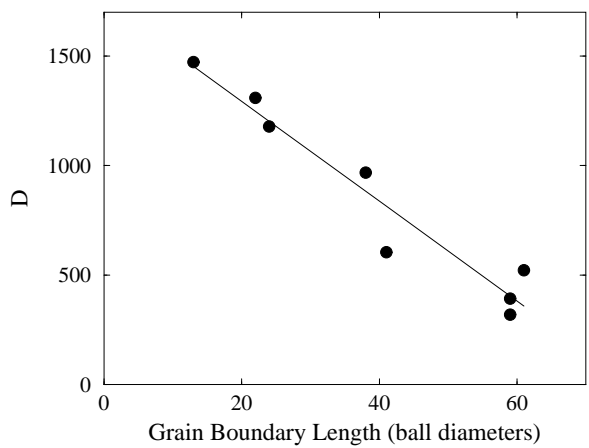}}
\caption{$D$, a ratio of correlation function values described in
the text, as a function of boundary length, for lattices perfect except for a
single grain boundary.
The line is a least-squares fit.}
\label{f:grainbdry}
\end{figure}

Values for best, worst, and initial arrangements appear in Table
\ref{t:peakratio}.  This probe is very sensitive to grain boundaries.  The two
most ordered runs of doubles have a single crystal spanning the entire
container. The third run has two regions with a grain boundary between them,
which greatly reduces the peak-to-valley ratio.  The best hexagon run has one
large domain and a small second domain, while the other runs have several small
domains in addition to the large one.  In fact, as shown in Figure 
\ref{f:grainbdry}, for systems with a single domain boundary the peak-to-valley
ratio is proportional to the length of the boundary.  With more than two
domains this linear relationship fails, but the high sensitivity of the
peak-to-valley ratio to boundaries remains.

Voids, interstitials, and other imperfections generally have much less effect
than boundaries.  One exception is that in nearly perfect lattices, the
first few valleys drop so close to zero that any disorder changes their
relative levels substantially; this causes the wide variation in the numbers
shown for doubles. A second exception is that occasionally, particularly
with doubles, a square lattice
forms in a small area.  This produces a peak on the histogram at $\sqrt{2}$
diameters, which is close to the center of the first valley and can have a
particularly large effect.

Doubles and hexagons show the best ordering, as they do by the long-range 
measures.  Once again, the two sizes of triangles are very similar in both the
best and worst order displayed.  One of the clearest indications of the
sensitivity of this index to grain boundaries is the noticeable difference
between diamonds and triples. The two shapes have comparable densities once
voids are filled in, and have visually similar amounts of ordering. However the
triples definitely have more grain boundaries, with fewer interstitials, which
is reflected here.

Properly normalizing for the order inherent in the different shapes is a
difficult problem.  Resolving the effect of internal correlations among the
balls composing a shape is particularly difficult with this measurement.  The
initial ordering corresponds well to cluster size, suggesting that the internal
correlations are important here.  However, this correlation disappears
after shaking, even for the least effective hit parameters.  We emphasize the
similarities between the two triangle shapes as evidence that this factor is
unimportant in the results from the final configurations.

As a complementary local order indicator, we calculate the Voronoi region of
each ball.  The Voronoi region of a ball consists of the points closer to that
ball than to any other.  The edges of the region are perpendicular bisectors of
the lines connecting the ball to its neighbors, so for a perfect triangular
lattice every Voronoi region would be a regular hexagon.  Pentagonal or
heptagonal regions signify defects in the lattice.  To measure order, we find
the percentage of balls with six-sided Voronoi regions.  This tells us how many
balls are in the interior of some domain.  

The highest percentage of hexagonal Voronoi regions ranges from over 98\% for
doubles to 78\% for trapezoids.  The highest, lowest, and initial percentages
for each shape appear in Table
\ref{t:Voronoi}.  These three numbers differ by at most 3.5\% for a single
shape, and usually by less than 1\%.  For comparison, the values for a typical 
initial configuration and for the most disordered final arrangements are also
shown.  Doubles order the most easily.  Hexagons achieve the next highest
order, although at the poorest hit settings they remain more disordered than
triangles.  Of the remaining shapes, triples and
diamonds have six nearest neighbors with about the same frequency.  Trapezoids
have significantly more irregular Voronoi regions.

We next address the issue of correlations within the larger shapes. The initial
configurations given in Table \ref{t:Voronoi} show no trend of higher order for
larger clusters.  The differences among the final configurations also exceed
correlation effects.  For example, each hexagon has a center ball with six
nearest neighbors, and centers account for 14.3\% of all balls.  For the three
most ordered hexagon configurations, on average 94.5\% of the balls have six
nearest neighbors, so 5.5\% do not.  Since all the center balls do, 6.4\% of
the non-centers must have imperfect Voronoi regions, a change of less than 1\%
in the number with hexagonal Voronoi regions.
The difference between hexagons and triangles is several times this large,
so the perfect arrangement around the center balls cannot explain the entire
difference.  The more complicated effects involve the outer balls.  For a
hexagon, each of the six outer balls has three nearest neighbors perfectly
positioned. A large triangle has three balls with four nearest neighbors, and
three others with two.  Without determining precisely the implications for the
Voronoi region's shape, we posit that the effect should be much smaller than
that of the central ball of the hexagon, and that we can reasonably ignore it
in distinguishing among shapes.  Also note that the hexagons form a very poorly
ordered initial state, despite the automatically proper coordination number of
the center balls.

\begin{figure}[b]
\includegraphics{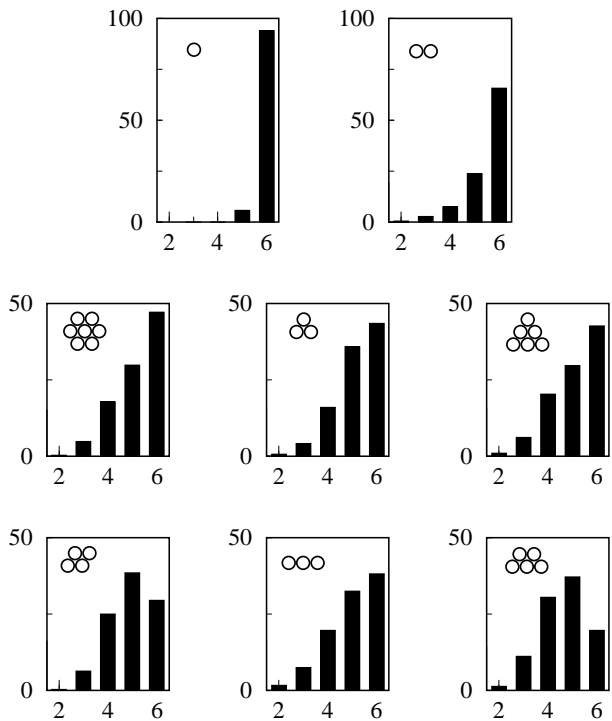}
\caption{Percentage of balls with a given coordination number, for the
configurations of Figure \ref{f:endconfigs}.}
\label{f:coord}
\end{figure}

The percentage of hexagonal Voronoi regions is particularly sensitive  to
holes.  A single missing ball in a perfect lattice creates six pentagonal
Voronoi regions, with larger holes disturbing more Voronoi regions.  Slight
lattice imperfections, such as those introduced by the ball center
identification, usually reduce the number of non-hexagonal Voronoi regions
around a single void to four.  Grain boundaries have less effect, with typically
one imperfect Voronoi region per ball of boundary length.  Interstitials have
little effect beyond the balls that compose them, since the balls at the
surface of a grain already have irregular Voronoi regions.

\begin{figure}[tbh]
\includegraphics{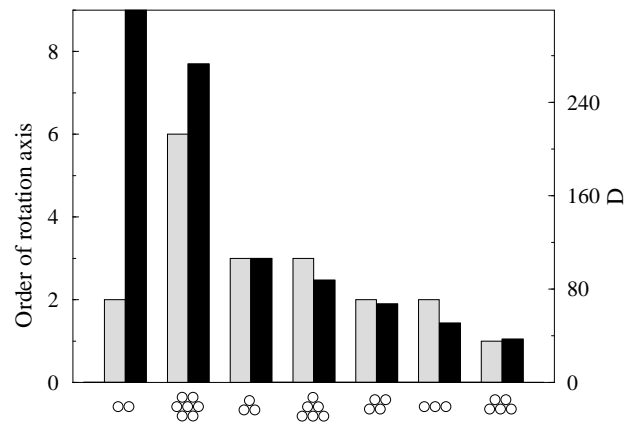}
\caption{A comparison of the average of the highest values $D$ (right) and
the number of symmetries (left) for each shape.  For doubles $D$ is actually
1191, far off the scale of this graph.}
\label{f:shapesym}
\end{figure}

The coordination number, the number of balls touching a given ball, is an
indicator related  but not identical to the Voronoi region shape.  No
coordination number can exceed six, although a Voronoi region can have more
than six sides.  Coordination numbers are also far more sensitive than Voronoi 
regions to small displacements of balls, for example from noise in identifying
the ball centers.  In Figure \ref{f:coord} we show the distribution of
coordination numbers for the configurations of Figure \ref{f:endconfigs}.  For
random sphere arrangements in three dimensions, the coordination number
distribution varies more than other properties \cite{3Dexpt}, including the
pair correlation function.  Balls are taken as touching when their
computer-identified centers lie within 1.08 ball diameters.  We choose this as a
cutoff because at longer distances significant asymmetry appears in the first
peak of the two-ball correlation function.  

Earlier two-dimensional experiments found a sharp change in average
coordination number near the random-ordered transition \cite{QT}. Although our
coordination numbers are larger due to correlations introduced by the welds,
they decrease abruptly near the same density.   The similar behavior gives
further encouragement to the possibility of studying random arrangements in two
dimensions.   

\section{Rotational symmetry}

We find that rotational symmetry is an excellent guide for predicting the
degree of order that a shape supports.  Figure \ref{f:shapesym} shows
both the average of the three highest peak-to-valley ratios $D$ 
(Table \ref{t:peakratio}) for each shape, and
also the order of the rotational axis.  Except for doubles, the
correspondence is excellent.  The various other measures of order
would give similar results.  The correspondence between ability to order and
rotational symmetry suggests that domains grow around the edges.  To
join the ordered portion, a shape must have the correct orientation. 
Its rotational symmetry sets a limit on the maximum angle through which
it needs to rotate to reach this orientation.  On the other hand,
the rotational symmetry has little correlation to the degree of order in
the initial configurations, which do not depend on growth at domain edges.

\begin{figure}[hbt]
\includegraphics{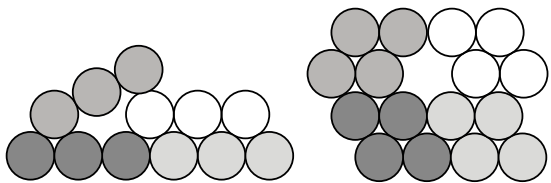}
\caption{Possible lattice growth with triples (left) and diamonds (right).}
\label{f:edgegrow}
\end{figure}

Several features of the final packings make sense when considering domain
growth. For some shapes, any stable position near a growing crystal is part of
the lattice.  For example, stability for doubles under gravity requires at
least three contacts, not all on the same ball.  At the edge of a crystalline
region, only lattice sites satisfy this criterion.  By contrast, triples
do have a stable off-lattice location, as illustrated in Figure
\ref{f:edgegrow}.  Trapezoids and large triangles also have stable nonlattice
positions for a single particle added to an existing lattice, but our other
shapes do not.  This may account for the significantly longer length scale of
small triangles compared to large ones. For diamonds, also illustrated in
Figure  \ref{f:edgegrow}, clusters positioned incompatibly may lead to voids,
explaining the large number of  voids in diamonds (see Table \ref{t:density}).
Finally, doubles can generally join an ordered region in any of three distinct
orientations, allowing them to order easily despite their lack of much
rotational symmetry.

\section{Conclusions and further work}

We have studied how several shapes composed of welded spheres pack in
two dimensions.  Our goals were to understand how particle shape affects
packings, and especially to identify two-dimensional random close-packed
configurations.  Constructing shapes as sphere clusters avoids issues
of maximum density and interparticle friction changing with shape, and
permits detailed comparison of the configurations reached by different shapes.
One consequence is that we find a strong correlation between the 
rotational symmetry of a particle and its short-range order.
In addition, the defects in long-range order relate to geometry.
Shapes where at least one side is three balls long produce more grain
boundaries, while voids are common with large nonlinear shapes.

In keeping with the tendency of shapes to order in two dimensions, most
of our clusters can anneal into configurations with long-range order
of the individual spheres.  Even under conditions where no long-range
order appears, the shapes form small domains with sharp boundaries,
qualitatively different from the random configurations that appear
in three dimensions.  Trapezoids are the one exception.  All their
packings remain random or nearly so, with typical domain sizes of only
two trapezoids.  Interestingly, the onset of ordered domains occurs near
the packing density 0.8, where previous experiments on disks already
showed evidence for a transition between random and ordered states.

Finding shapes such as trapezoids with stable random arrangements in two 
dimensions allows comparison to the packing behavior of spheres in three
dimensions. We plan to pursue the similarities further by studying the time
dependence of trapezoid configurations during annealing.  The data on
rotational symmetry suggest that examining elongated shapes with little
symmetry may identify other shapes that do not crystallize.  However, because
of difficulties with breakage and with system size, this work is better done
through computer simulations.  The data presented here provide a series of test
cases on the realism of any simulations.

Although our artificial particles are convenient for comparisons among the
shapes, the unusual surface geometry clearly changes some behavior from that of
similar but convex shapes.  The irregular surfaces allow neighboring particles
to lock together, leading to the high void densities for our larger shapes.
More significantly, only the constituent spheres form ordered structures, {\em
not} the larger shapes.  For example, two-dimensional simulations of prolate
ellipses under the influence of gravity find orientational but not
translational long-range order \cite{Buchalter2D}.  By contrast, our doubles
have no long-range orientational order.  Instead, the dimples in the shapes'
sides allow neighboring doubles to interlock and overcome the effect of
gravity. A further project would deform doubles gradually into ellipsoids by
filling in the dimples, while tracking changes in the characteristic
arrangements.  Once again, this project is most practical through simulations.

We are continuing work along other lines as well.  Time
dependence measurements for shapes that do crystallize may help in 
understanding how non-spherical particles move into position.  Finally, we
plan to extend the measurements towards three dimensions by varying the
container thickness to accommodate more layers of balls.

\section{Acknowledgements}

We thank J.D. Lawton for help in setting up the apparatus.  This work is
supported by the National Science Foundation under DMR-9733898.

\end{document}